\documentclass[aps,twocolumn,showpacs,amssymb,pra]{revtex4-1}
\usepackage{amsmath}
\usepackage{graphicx}
\usepackage{dcolumn}
\usepackage{bbm}
\usepackage{color}

\begin{document}
\title{One-dimensional Bose--Hubbard model 
with local three-body interactions}

\author{Satoshi Ejima}
\author{Florian Lange}
\author{Holger Fehske}
\affiliation{Institut f{\"u}r Physik, Ernst-Moritz--Arndt-Universit{\"a}t Greifswald, D-17489 Greifswald, Germany}
\author{Florian Gebhard}
\author{Kevin zu M{\"u}nster}
\affiliation{Fachbereich Physik, Philipps Universit{\"a}t Marburg, D-35032 Marburg, Germany}

\begin{abstract}%
We employ the (dynamical) density matrix renormalization group technique 
to investigate the ground-state properties of the Bose--Hubbard
model with nearest-neighbor transfer amplitudes~$t$
and local two-body and three-body repulsion
of strength~$U$ and $W$, respectively.
We determine the phase boundaries between the Mott-insulating
and superfluid phases for the lowest two Mott lobes from the
chemical potentials. We calculate the tips of the Mott lobes from
the Tomonaga--Luttinger liquid parameter and confirm the
positions of the Kosterlitz--Thouless points from the 
von Neumann entanglement entropy. 
We find that physical quantities in the second Mott lobe such as
the gap and the dynamical structure factor 
scale almost perfectly in $t/(U+W)$, even close to the Mott transition.
Strong-coupling perturbation theory shows that there is no true scaling 
but deviations from it are quantitatively small in the strong-coupling limit.
This observation should remain true in higher dimensions and for not too
large attractive three-body interactions.
\end{abstract}


\maketitle

\section{Introduction} 
The revolutionary control over ultracold atoms in optical lattices
made possible the direct experimental 
observation of many-body states of different quantum systems~\cite{BDZ08}. 
Tuning two-body interactions
by using Feshbach resonances or changing the strength of the lattice 
confinement permitted the observation of a superfluid (SF) to
Mott-insulating (MI) quantum phase transition for bosonic lattice atoms~\cite{GMEHB02}
for integer densities.
More recently, the importance of
multi-body interactions has been inferred from
experiment~\cite{WBSHLB10,MHLDDN11}. These interactions
were so far deemed negligible 
higher-order many-body interactions.

The minimal model to describe bosonic lattice quantum gases 
is the Bose--Hubbard model. On a chain with $L$ sites,
the Hamilton operator with local two-body and three-body interactions 
is given by
\begin{eqnarray}
 H &=& -t\hat{T}+U \hat{D}+W \hat{W}\; , \nonumber \\
\hat{T}&=& \sum_{j=1}^{L-1} 
         \left(\hat{b}_j^\dagger \hat{b}_{j+1}^{\phantom{\dagger}}
+ \hat{b}_{j+1}^\dagger \hat{b}_{j}^{\phantom{\dagger}}\right)\; , \nonumber \\
\hat{D} &=& \frac{1}{2}\sum_{j=1}^L \hat{n}_j(\hat{n}_j-1) \; , 
      \nonumber \\
\hat{W}&=& \frac{1}{6}\sum_{j=1}^L \hat{n}_j(\hat{n}_j-1)(\hat{n}_j-2)\; , 
 \label{Hamil}
\end{eqnarray}
where $\hat{b}_j^\dagger$ and $\hat{b}_j^{\phantom{\dagger}}$
are the creation and annihilation operators for bosons on site $j$,
$\hat{n}_j=\hat{b}_j^\dagger\hat{b}_j^{\phantom{\dagger}}$ 
is the boson number operator on site $j$, $t$ is the tunnel amplitude
between neighboring lattice sites, and $U>0$ ($W>0$) denote the strength 
of the on-site two-body (three-body) repulsion. 
There are $N=\rho L$ bosons in the system. 

The Bose--Hubbard model in one dimension with pure two-body interactions
has been extensively studied~\cite{KM98,KWM00,EFG11,EFGMKAL12},
using the density matrix renormalization group (DMRG) technique~\cite{Wh92}.
SF-to-MI quantum phase transition points can be determined accurately
from the finite-size scaling of the Tomonaga--Luttinger liquid (TLL) parameter $K_b$;
the model is in the universality class of the spin-1/2 $XY$ 
model so that there exists a Kosterlitz-Thouless transition
in the thermodynamic limit, $L\to\infty$, where $K_b=1/2$.

The model including the three-body
local interaction remains largely unexplored. The first Mott lobe
was recently studied using DMRG~\cite{SS11}
and the cluster expansion approach~\cite{VM12}.
The whole ground-state phase diagram 
for the first and second lobes was derived from exact diagonalization results
for small system sizes $L\leq 8$~\cite{So12} 
and DMRG data for larger system sizes~\cite{SDMPD12}.
Very recently, Sowi{\'n}sky et al.~\cite{SCDTL13} studied the full model 
with attractive two-body and repulsive three-body interactions
using the DMRG.
 
In this work, we examine 
the static and dynamical ground-state properties of the Bose--Hubbard model 
with two-body and three-body interactions 
using large-scale DMRG calculations.
To determine the phase boundaries between MI and SF phases 
we compute the chemical potential, the Tomonaga--Luttinger liquid (TLL) 
parameter, and the von Neumann entanglement entropy. 
Moreover, for $\rho=2$, we calculate the single-particle gap 
and the dynamical structure factor $S(q,\omega)$ 
using the dynamical DMRG (DDMRG) procedure~\cite{Je02b}. 

We find that the three-body interaction is qualitatively irrelevant in one
dimension for strong coupling. Its quantitative corrections
for $\rho=2$ for the size of the Mott lobe, the gap, and the 
dynamical structure factor can be incorporated almost perfectly 
by rescaling $t/U\to t/(U+W)$ in the results for a bare two-body interaction ($W=0$).
Strong-coupling perturbation theory shows that this rescaling is not
rigorous but quantitative corrections are very small 
for strong interactions. 

\begin{figure*}[t]
\begin{center}
\includegraphics[width=0.9\linewidth,clip]{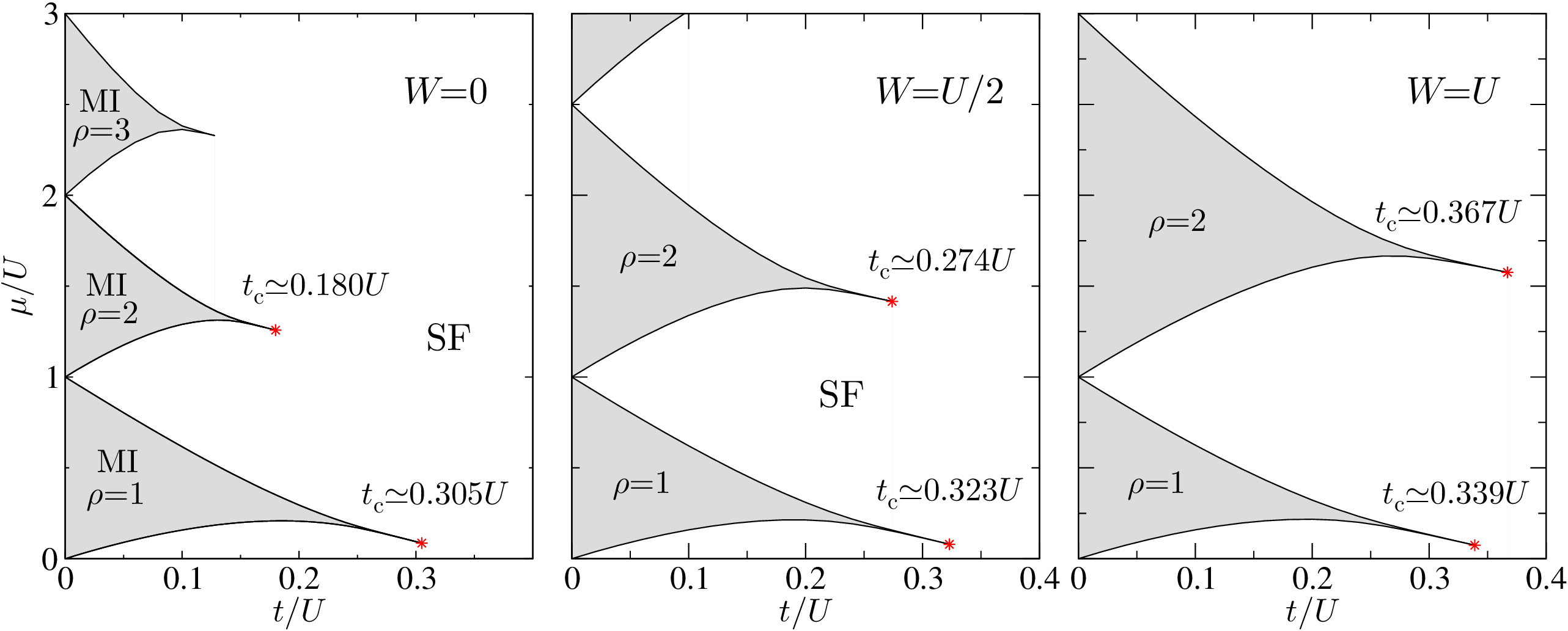}
\end{center}
\caption{(Color online) Ground-state 
phase diagram of the one-dimensional Bose--Hubbard model
with three-body interactions for $W=0$ (left panel), 
$W=U/2$ (middle panel), and $W=U$ (right panel), 
showing superfluid (SF) and Mott insulating (MI) regions
as a function of the chemical potential~$\mu/U$
and the electron transfer amplitude~$t/U$. Results are based on DMRG data 
for lattices with up to  $L=128$ sites and open boundary conditions.
The position of the Mott tips is obtained from
the finite-size extrapolation 
of the Tomonaga--Luttinger parameter $K_b(L)$, 
eq.~(\protect\ref{eq:Kb}).\label{fig:pd}}
\end{figure*}

\section{Ground-state phase diagram}

\subsection{Single-particle gap}
\label{subsec: spgap}

The Mott-insulating phases are characterized by a finite gap for single-particle
excitations. The chemical potential $\mu(L,N)$ gives the energy
for adding the $N$th particle to the system with $L$ sites in its ground state,
\begin{equation}
\mu(L,N)=E_0(L,N)-E_0(L,N-1) \; ,
\label{eq:chempotential}
\end{equation}
where $E_0(L,N)$ is the ground-state energy for $N$ bosons on $L$~sites.
For $t=0$, $[\mu(L,L+1)=U, \mu(L,L)=0]$ and
$[\mu(L,2L+1)=2U+W, \mu(L,2L)=U]$ hold.
The single-particle gap results from 
the difference between the chemical potentials
for $N+1$ and $N$ particles,
\begin{equation}
\Delta(L,N)=\mu(L,N+1)-\mu(L,N) \;.
\label{eq:gapdef}
\end{equation}
In the Mott insulating state, the gap remains finite in the thermodynamic limit,
$\Delta(\rho)\equiv \Delta(L\to\infty,N\to\infty)>0$ for fixed $\rho=N/L$.
In the superfluid phase, the chemical potential is a continuous
function of the density~$\rho$.
For $t=0$, we have 
$\Delta(\rho=1)=U$  and $\Delta(\rho=2)=U+W$.

For integer fillings~$\rho$ and for large $U/t$, the Bose--Hubbard model
is in its Mott insulating phase. Thus, the ground-state phase diagram 
in the plane $(t/U,\mu/U)$ shows superfluid and Mott-insulating regions,
see Fig.~\ref{fig:pd}.
In the shaded regions, the particle density is constant and the 
gap for fixed $t/U$ is given by the difference in the chemical potentials 
at the boundaries to the superfluid phases, i.e.,
the chemical potentials determine the phase boundaries.
Unfortunately, the gaps become exponentially small close to the
critical points $t_{\rm c}^{\rho}$ above which the Mott gap is zero. 
Since it is not possible to resolve such small gaps numerically,
the critical points must be determined in a different manner, 
see Sect.~\ref{subsec:KTpoints}.
We obtained the ground-state phase diagram in Fig.~\ref{fig:pd} 
using the DMRG method with up to $L=128$ sites and 
open boundary conditions (OBCs).
We extrapolated the chemical potentials to the thermodynamical limit
using a second-order polynomial fit in $1/L$ for $\mu(L,L+1)$
and $\mu(L,L)$~\cite{EFG11,EFGMKAL12}.

In the absence of the three-body interaction (left panel),
the Mott lobes become smaller with increasing density $\rho$,
and, concomitantly, the values of the transition points $t_{\rm c}^{\rho}$
decrease with increasing $\rho$.
When we switch on the three-body interaction (middle panel for $W=U/2$,
right panel for $U=W$),
the first Mott lobe is almost unchanged in comparison with the result for
$W=0$, and the dependence of $t_{\rm c}^{\rho=1}$ on~$W$ is 
rather weak, too. This is readily understood because
at $t_{\rm c}^{\rho=1}\simeq 0.3 U$, there are few doubly occupied sites 
in the ground state and hardly any triply occupied sites: the Mott transition
occurs at strong coupling where triple occupancies are strongly reduced
for $\rho=1$.
The situations is different for $N=2 L$ particles because
there are essentially double and triple occupancies present at the transition
so that the three-body interaction is effective. 
Therefore, the size of the second Mott lobe and 
the value of $t_{\rm c}^{\rho=2}$ substantially increase as a function of~$W$,
which also pushes up in energy the other Mott lobes.

\subsection{Critical interactions}
\label{subsec:KTpoints}

In order to determine the transition points, 
we compute the TLL parameter $K_b$ in the superfluid phase. 
The Fourier transformation of the static structure factor,
$S(q)=(1/L)\sum_{j,l}e^{{\rm i}q(j-l)}(\langle\hat{n}_j\hat{n}_l\rangle-
                 \langle\hat{n}_j\rangle\langle\hat{n}_l\rangle)$ 
at $q=2\pi/L$ defines $K_b(L)$~\cite{EGN05,EFG11},
\begin{equation}
 \frac{1}{2\pi K_b(L)}=\frac{S(2\pi/L)}{2\pi/L}\; .
\label{eq:Kb}
\end{equation}
The TLL exponent is then obtained from the extrapolation of the
DMRG data,  $K_b=\lim_{L\to\infty}K_b(L)$.
At the SF-MI transition point we expect that $K_b=1/2$~\cite{Gi03}, 
as in the case for the model (\ref{Hamil}) at $W=0$~\cite{KM98,KWM00,EFG11}.
In this way, the transition points $t_{\rm c}^{\rho}$ can be determined accurately.

\begin{figure}[tb]
\begin{center}
\includegraphics[width=\columnwidth,clip]{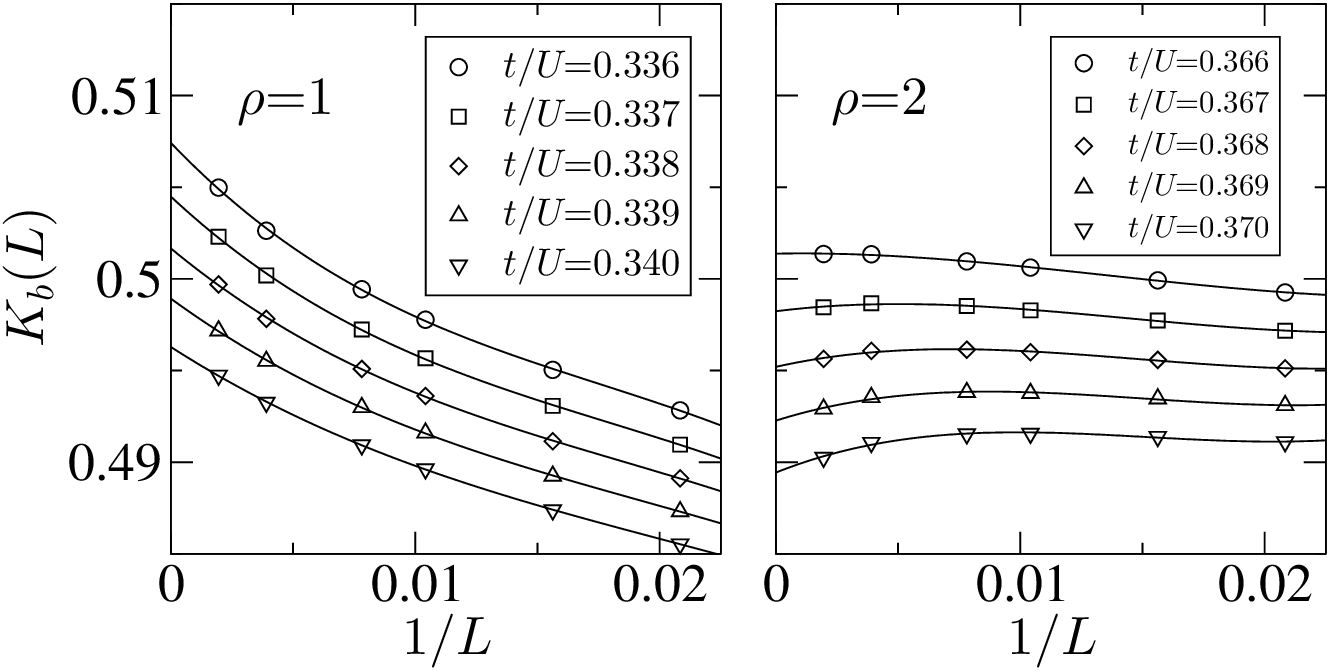}
\includegraphics[width=\columnwidth,clip]{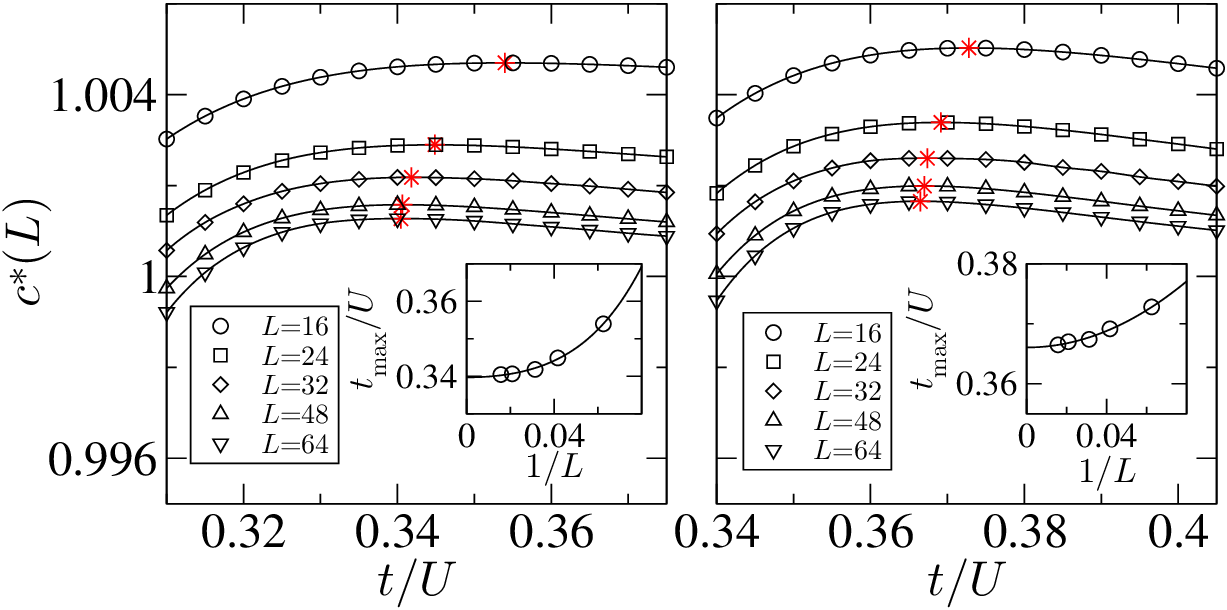}
\end{center}
\caption{(Color online) DMRG data for the 
 Tomonaga--Luttinger liquid parameter $K_b(L)$ (upper panels) 
 and the central charge $c^{\ast}(L)$ (lower panels)
 for the first (left panels) and second (right panels)
 Mott lobes in the model~(\protect\ref{Hamil}) for $W/U=1$.
 The position of the maxima of $c^\ast(L)$ can be  extrapolated 
 systematically to the thermodynamic limit,
 as shown in the insets of the lower panels.\label{fig:Kb-cc} 
}
\end{figure}

In the upper panels of Fig.~\ref{fig:Kb-cc} we demonstrate
the finite-size scaling of the TLL parameter $K_b(L)$ 
for $\rho=1$ (left panel) and $\rho=2$ (right panel) at $U=W$,
where we use the DMRG method with up to $L=512$ sites and OBC. 
The transition points are determined from $\lim_{L\to \infty}K_b(L)=0.5$
as $t_{\rm c}^{\rho=1}=0.339\pm0.001$ and 
$t_{\rm c}^{\rho=2}=0.367\pm0.001$. Our data improve the estimates
from the exact diagonalization method~\cite{So12}
with up to $L=8$ sites,
$t_{{\rm c},{\rm ED}}^{\rho=1}/U= 0.32$ and $t_{{\rm c},{\rm ED}}^{\rho=2}/U=0.38$.

In order to confirm the values of the critical points,
we address the von Neumann entanglement entropy $S_L(\ell)$,
defined as $S_L(\ell)=-{\rm Tr} _{\ell}(\rho_\ell\ln\rho_\ell)$ with the 
reduced density matrix $\rho_\ell={\rm Tr}_{L-\ell}(\rho)$.
For a system with central charge $c$,
one finds for periodic boundary conditions (PBCs)~\cite{CC04} 
\begin{equation}
  S_L(\ell)=\frac{c}{3}\ln\left[
             \frac{L}{\pi}\sin\left(\frac{\pi\ell}{L}\right)
           \right]
           +s_1\;,
\label{eq:von-neuman-entropy}
\end{equation} 
where $s_1$ is a non-universal constant and corrections are small, of the order~$1/L$.
The constant~$s_1$ can be eliminated by 
subtracting $S_L(L/2)$ from $S_L(L/2-1)$ 
so that we can define the size-dependent central charge~\cite{Ni11} 
\begin{equation}
  c^*(L)=
\frac{3\left[S_L(L/2-1)-S_L(L/2)\right]}{\ln\left[\cos(\pi/L)\right]}\;.
\label{eq:cstar}
\end{equation}
Since the excitations of the interacting suprafluid in one dimension 
form a TLL with central charge $c=1$,
we can use $c^*=\lim_{L\to\infty}c^*(L)$ to locate the critical interactions.

In the lower panels of Fig.~\ref{fig:Kb-cc}, we employ the DMRG method 
for up to $L=64$ sites and PBCs. 
As in the case of the model (\ref{Hamil}) for $W/U=0$~\cite{EFGMKAL12}, 
$c^{\ast}(L)$ displays maxima as a function of 
the hopping amplitude $t/U$ for fixed system size $L$.
When we extrapolate the position of the maxima in $c^\ast(L)$ for $U=W$
to the thermodynamic limit, we obtain the transition points 
as $t_{\rm c}^{\rho=1}/U=0.340\pm 0.002$ and 
$t_{\rm c}^{\rho=2}/U\simeq 0.366\pm 0.002$, in excellent agreement
with our previous estimate from the TLL parameter $K_b$. 
In this way, we reliably determined the position of the tips
in the Mott lobes of the ground-state phase diagram in Fig.~\ref{fig:pd}.

\section{Second Mott lobe}

In the remainder of this work we focus on the second Mott lobe,
$\rho=2$ and $t<t_{\rm c}^{\rho=2}$, where the three-body repulsion plays a significant role. 

\subsection{Gap function}

\begin{figure}[tb]
\begin{center}
\includegraphics[clip,width=\columnwidth]{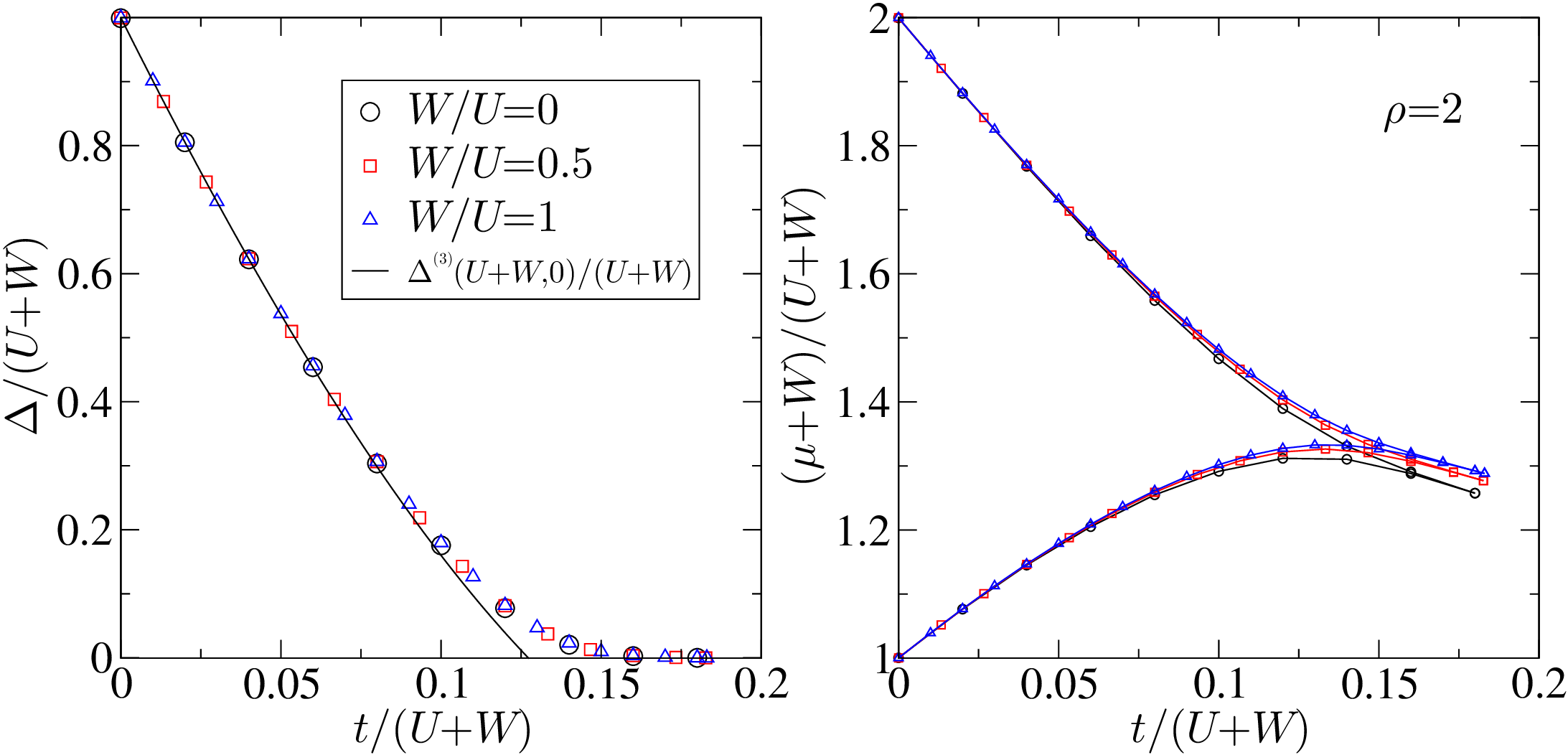}
\end{center}
\caption{(Color online) Rescaled single-particle gaps (left panel) and
phase diagram (right panel) 
of the second Mott lobe ($\rho=2$) in the model~(\protect\ref{Hamil}).
The data for $W=0,U/2,U$ collapse onto each other.
Also shown is the third-order result for the gap $\Delta^{(3)}(U+W,0)$
at $U=W$, eq.~(\protect\ref{eq:gap-PT}).\label{fig:Delta}
}
\end{figure}

For strong interactions, the gap can be calculated systematically
in a power series in the particle hopping. For $N=2L+1$ particles,
the first non-trivial order describes the free propagation 
of a triple occupancy so that its dispersion relation is given by
$\omega^+(k)=E^+(k)-E_0(L,2L)=W+2U -6t\cos(k)$.
For $N=2L-1$ particles, the singly occupied site (a hole in the 
background of doubly occupied sites) has the dispersion relation
$\omega^-(k)=E^-(k)-E_0(L,2L)=-U +4t\cos(k)$.
Therefore, to leading order in $t$, the gap becomes
$\Delta={\rm Max}_k[\omega^+(k)-\omega^-(k)]$ or
\begin{equation}
\frac{\Delta(t\ll U,W)}{U+W}=1-\frac{10 t}{U+W} + \ldots
\end{equation}
To leading order, $\Delta$ is only a function of $t/(U+W)$.
In the left panel of Fig.~\ref{fig:Delta} we 
plot $\Delta/(U+W)$ as a function of $t/(U+W)$ for $t<t_{\rm c}^{\rho=2}$.
It is seen that the data for $W=0,U/2,U$ essentially
collapse onto a single curve, suggesting that the gap is solely
a function of $U+W$. Likewise, when we plot the modified chemical
potentials $\mu+W$ as a function of $t/(U+W)$ in the right panel of
Fig.~\ref{fig:Delta}, we see that the second Mott lobes 
for $W=0,U/2,U$ almost collapse onto each other.
Deviations are visible only close to the transition points.

\subsection{Dynamical structure factor}

The dynamical structure factor $S(q,\omega)$ characterizes 
the density-density response of the Bose gas and is directly accessible 
by Bragg spectroscopy~\cite{SICSPK99,EGKPLPS09}.
It is defined as
\begin{equation}
S(q,\omega)
=\sum_n|\langle\psi_n|\hat{n}(q)|\psi_0\rangle|^2\delta(\omega-\omega_n),
\end{equation} 
where $\hat{n}(q)=\sum_le^{{\rm i}ql}\hat{b}_l^\dagger\hat{b}_l^{\vphantom{\dagger}}$,
$|\psi_0\rangle$ and $|\psi_n\rangle$ denote the ground state and $n$th 
excited state, respectively, and $\omega_n=E_n-E_0$ gives 
the corresponding excitation energy.
As shown in Ref.~\cite{EFG12,EFGMKAL12}, we can compute 
the Lorentz-broadened
$S_{\eta}(q,\omega)$ with
\begin{equation}
S_{\eta}(q,\omega)= \int_{-\infty}^{\infty}{\rm d} \omega' S(q,\omega')
\frac{\eta}{\pi[(\omega-\omega')^2+\eta^2]}
\end{equation}
for Bose--Hubbard models using the DDMRG technique.
In this work, we use $\eta=2t$ for $L=32$ sites with PBCs
to obtain smooth curves.
Note that the unbroadened dynamical quantities corresponding 
to the experimental measurements can be extracted from the
deconvolution of the DDMRG data as demonstrated in Ref.~\cite{NGJ04}. 
\begin{figure}[ht]
\begin{center}
\includegraphics[clip,width=\columnwidth]{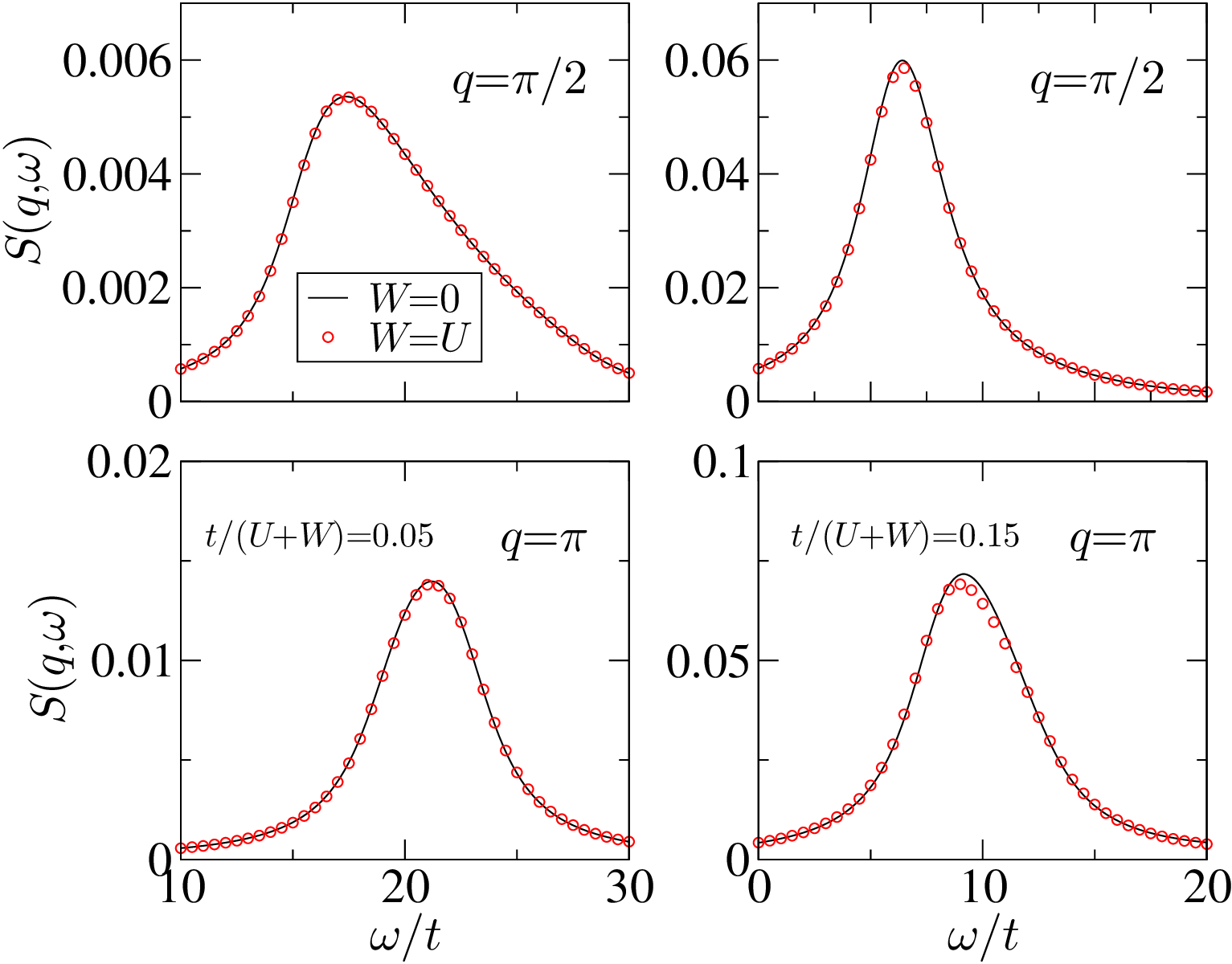}
\end{center}
\caption{(Color online) 
Dynamical structure factor $S_{\eta}(q,\omega)$ with Lorentz broadening
$\eta=2t$ at momenta $q=\pi/2$ (upper panels) 
and $q=\pi$ (lower panels) of the second Mott 
lobe in the model~(\protect\ref{Hamil})
for $t/(U+W)=0.05$ (left panels) 
and $t/(U+W)=0.15$ (right panels), as obtained from  
the DDMRG technique for $L=32$ sites.
\label{fig:Skw}
}
\end{figure}

In Fig.~\ref{fig:Skw} we show $S_{\eta}(q,\omega)$ for $q=\pi/2$ 
(upper panels) and $q=\pi$ (lower panels)
for $t/(U+W)=0.05$ (left panels) and $t/(U+W)=0.15$ (right panels)
as a function of $\omega/t$. The results for $W=U$ lie essentially on top
of those for $W=0$, apart from small deviations around $\omega=0$
for $t/(U+W)=0.15$ and $q=\pi$, suggesting again a scaling relation.

\subsection{Strong-coupling perturbation theory}
\label{subsec:analysics}

The strong-coupling analysis of Harris and Lange as used in~\cite{EFGMKAL12}
starts from the unperturbed Hamiltonian $\hat{H}_0=U\hat{D}$
where $\hat{D}$ counts  the number~$n_D$ of
pair interactions. The perturbation expansion for $t/U\to 0$
relies on the fact that the subspaces with $n_D=0,1,2,\ldots$ are well separated.
In the presence of the triple interaction~$\hat{W}$, we assume energetically
separated subspaces 
for $\hat{H}_0=U\hat{D}+W\hat{W}$ that have the unperturbed energies
$E_0(n_D,n_W)=Un_D+Wn_W$ with two integer quantum numbers 
$(n_D\geq 0, n_W\geq 0)$. 
A transfer process between neighbors with positive integer occupancies 
$(n_1=\rho+m,n_2=\rho+n)$ to $(n_1=\rho+m-1,n_2=\rho+n+1)$ 
results in an energy transfer
\begin{equation}
\Delta E(m,n)=(n-m+1)\left[U+W(\rho-1) +W(n+m)/2\right] \; .
\end{equation}
For $\rho=2$, the resulting finite 
energy denominators in a perturbation expansion
thus include the energies $U+W+W(n+m)/2$. 
Therefore, there cannot be a 
rigorous scaling relation, i.e., the physical quantities are
not solely a function of $t/(U+W)$.

However, the lowest-order terms involve
small deviations from $\rho=2$, i.e., only 
$(m=0,n=0)$ and $(m=\pm1, n=\mp 1)$ appear.
Thus, only the energy denominator $U+W$ occurs
in low-order perturbation theory in $\hat{T}$. Moreover,
in higher orders, those terms that involve
$t/(U+W)$ have a higher weight
than those that are proportional
to $t/(2U+W)$ or else. For these practical reasons, the results 
for the gap and the dynamical structure factor 
show fairly small deviations from a scaling behavior, apart from
the region close to the MI-to-SF transition.

\subsubsection{Gap}

To be definite, we address the series expansion of the gap
to third order. To this end, we repeat the perturbative 
expansion~\cite{EFGMKAL12,MuensterMSc} whereby we effectively replace 
the energy denominators $U(n_{D,1}-n_{D,2})$ of states
with $n_{D_1}$ and $n_{D_2}$ double occupancies by the
appropriate energy differences 
$U(n_{D,1}-n_{D,2})+W(n_{W,1}-n_{W,2})$.
For $\rho=2$ this leads to
\begin{eqnarray}
\frac{\Delta^{(3)}(U,W)}{U+W}&=& 1 - \frac{10 t}{U+W}\nonumber \\
&&
+\frac{2t^2 \left(26 U^2+57 U W+19 W^2\right)}{(U+W) ^2(2 U+W) (2 U+3 W)} 
\\
&& +\frac{12t^3 }{(U+W)^3 (2 U+W)^2 (2 U+3 W)^2} \nonumber\\
&&\hphantom{+}[
40 U^4+160 U^3 W+230 U^2 W^2\nonumber \\ 
&&\hphantom{+ [ }+ 134 U W^3+29 W^4]\nonumber
 \; .
\label{eq:gap-PT}
\end{eqnarray}
It is seen that deviations from the scaling of $\Delta/(U+W)$
with $t/(U+W)$ first occur in second order. 
In order to assess the quantitative effect, 
we use the result for $W=0$ and define
\begin{eqnarray}
\frac{\Delta^{(3, {\rm sc})}(U,W)}{U+W}&=& \frac{\Delta^{(3)}(U+W,0)}{U+W} \\ 
&=& 1 - \frac{10 t}{U+W} + \frac{13t^2}{(U+W)^2}
+ \frac{30t^3}{(U+W)^3} \, .\nonumber 
\end{eqnarray}
Then, at $U=W$ we consider
\begin{eqnarray}
\delta^{(3)}(U) &=& 
\frac{|\Delta^{(3)}(U,U)-\Delta^{(3, {\rm sc})}(U,U)|}{\Delta^{(3)}(U,U)} 
\\
&=&
\frac{t^2 (61 t + 45 U)}{2 (593 t^3 + 510 t^2 U - 750 t U^2 + 150 U^3)} \; .
\nonumber
\end{eqnarray}
where, e.g., $\delta^{(3)}(t/0.15)=0.012$.
At $U=W$ we find that the relative deviation
of the scaling curve for $t/U< 0.15$
is about one percent or less, a negligibly small correction. This explains
the almost perfect scaling seen in Fig.~\ref{fig:Delta}.

\subsubsection{Dynamical structure factor}

As has been shown in Refs.~\cite{EFGMKAL12,MuensterMSc}, the dynamical
structure factor in the region of the primary Hubbard band
can be obtained from the solution of an effective
single-particle problem on a ring that describes the propagation 
of a `hole' (single occupancy for $\rho=2$)
and a `particle' (triple occupancy for $\rho=2$)
that move with total momentum $q$. The resulting effective
single-particle problem is governed by a kinetic term
and a potential whose range is proportional to the order of the expansion.

In the calculation of the structure factor, we need 
the eigenstates $|q;k\rangle$ 
of $\overline{k}_0$ that describes the free propagation of particles
and holes,
\begin{eqnarray}
\overline{k}_0 |q;k\rangle 
&=& (\beta_q e^{-{\rm i}k} + \beta_q^* e^{{\rm i}k} ) |q;k\rangle \; ,
\nonumber \\
\beta_q &=& -t[\rho +(\rho +1)e^{{\rm i} q}] \; .
\end{eqnarray}
The states $|q\rangle$ that enter the calculation of the structure factor
are weighted linear combinations of the states $|q;k\rangle$.
Up to second order, these weights are solely a function of $t/(U+W)$
and do not play a role for our discussion.

By definition, the states $|q;k\rangle$ are eigenstates of the kinetic contribution
to all orders,
\begin{eqnarray}
\overline{k}_1 |q;k\rangle 
&=& (\alpha + \gamma_q e^{-{\rm i}2k} + \gamma_q^* e^{{\rm i}2k} ) |q;k\rangle \; ,
\nonumber \\
\gamma_q &=& -t\frac{t\rho(\rho+1)}{U+(\rho-1)W}(1+e^{{\rm i} 2q}) \; ,\\
\alpha &=& t \frac{8t\rho(\rho+1)}{U+(\rho-1)W}
- t \frac{2t(\rho^2-1)}{2U+(2\rho-3)W} \nonumber \\
&& - t \frac{2t\rho(\rho+2)}{2U+(2\rho-1)W} \nonumber \; .
\end{eqnarray}
For $\rho=2$, $\alpha$ is not a function of $t/(U+W)$ alone but also
involves terms of different analytical structure.

\begin{figure}[tb]
\begin{center}
\includegraphics[clip,width=\columnwidth]{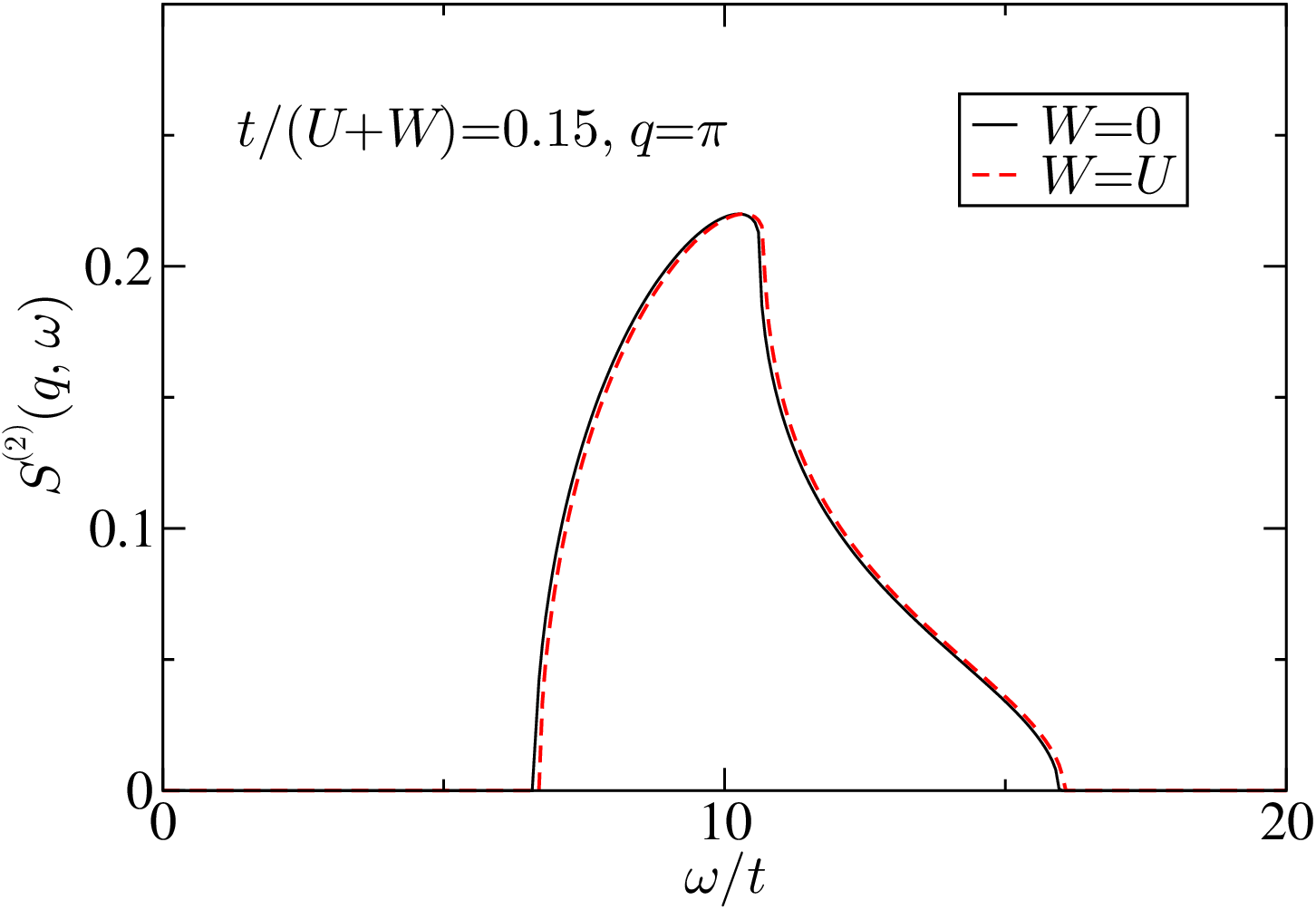}
\end{center}
\caption{(Color online)  Dynamical structure factor 
in second-order strong-coupling perturbation theory,
$S^{(2)}(q=\pi,\omega)$, for $U=W$ 
and $t/(U+W)=0.15$ (full line),
in comparison with the scaling result $S^{(2,{\rm sc})}(q=\pi,\omega)$
(dashed line).  The differences are very small. \label{fig:kevinfig}}
\end{figure}

The same observation holds for the interaction potential.
Apart from the hard-core correction that appears
when we close the chain into a ring~\cite{EFGMKAL12,MuensterMSc}, 
the interaction potential contains terms between nearest neighbors 
and next-nearest neighbors of the form
\begin{equation}
\frac{V_{1,1}(q)}{t}= \frac{2t\rho(\rho+1)\cos(q) 
+2t(2\rho+1)^2/3}{U+(\rho-1)W} \; ,
\end{equation}
and 
\begin{eqnarray}
\frac{V_{L-1,1}(q)}{t}&=& \frac{t\rho(\rho+1)}{U+(\rho-1)W}
- \frac{t\rho(\rho+2)}{2U+(2\rho-1)W} \nonumber \\
&& + \frac{t\rho(\rho+1)e^{-{\rm i}q}}{U+(\rho-1)W} \\
&& + \frac{t\rho(\rho+1)e^{-{\rm i}2q}}{U+(\rho-1)W}
- \frac{t(\rho-1)^2 e^{-{\rm i}2q}}{2U+(2\rho-3)W} \; . \nonumber
\end{eqnarray}
The contributions for next-nearest neighbors involve
intermediate states with excitation energies that are different
from $U+W$ or its multiples for $\rho=2$.
In Fig.~\ref{fig:kevinfig} we show the resulting structure
factor to second order $S^{(2)}(q,\omega)$ 
at $q=\pi$ as a function of $\omega$
for $U=W$, $t/U=0.15$, and $\rho=2$,
where we give the full result to second order. For comparison
we also show $S^{(2,{\rm sc})}(q,\omega)$ 
where we replace $U$ by $U+W$ in the results for $W=0$.
The quantitative differences are very small so that 
the scaling $U \to U+W$ is almost fulfilled, 
as seen in the DDMRG data. 
Note that the result of $S(q,\omega)$ in Fig.~\ref{fig:kevinfig}
looks quantitatively different from the DDMRG data
in the right lower panel of Fig.~\ref{fig:Skw}. This is due to the 
finite broadening width $\eta$ in the DDMRG calculation and the limited
validity of the second-order strong-coupling approach for intermediate
coupling strengths.

\section{Conclusions}

In this work we used the numerically exact 
density matrix renormalization group technique to
investigate the one-dimensional Bose--Hubbard model
with a two-body interaction~$U\hat{D}$ and a three-body interaction $W\hat{W}$.
We determined the ground-state phase diagram with and without three-body repulsion 
for the first and second Mott lobes from the chemical potentials.
The calculation of the Tomonaga-Luttinger parameter $K_b$ 
close to the Kosterlitz--Thouless transitions provided accurate estimates
for the tips of the Mott lobes where the gaps become exponentially small.
We confirmed the results for $t_{\rm c}^{\rho=1,2}$ 
using the von Neumann entanglement entropy.
We find that $t_{\rm c}^{\rho=1}(W)$ increases from $t_{\rm c}^{\rho=1}(0)$ by
less than ten percent for $W\leq U$. Moreover, 
$t_{\rm c}^{\rho=2}(W)$ obeys the relation 
$t_{\rm c}^{\rho=2}(W)\approx t_{\rm c}^{\rho=2}(0)(1+W/U)$
with an accuracy of a few percent for $W\leq U$.

For the second Mott lobe, $\rho=2$,
our numerical data for the gap $\Delta$
and the Lorentz-broadened dynamical structure factor $S_{\eta}(q,\omega)$
showed a similarly good `scaling behavior': the curves for different values 
of $W=0,U/2,U$ almost fall on top of each other when plotted as a function 
of $t/(U+W)$. Our results from strong-coupling perturbation theory show that
the scaling is not rigorous. However, deviations are quantitatively small because,
in the Mott insulating region, interactions are strong so that
the physical quantities are largely determined by particle-hole excitations with
one triple occupancy  and one single occupancy
in the background of doubly occupied sites.

Since our observations rely on the strong-coupling picture alone, they
should also hold in higher dimensions.
The Bose--Hubbard model with pure two-body interactions
remains qualitatively correct even in the presence of sizable
local three-particle interactions. Of course, this conclusion becomes invalid for 
substantial attractive three-body interactions 
that will lead to the formation of local clusters.

\acknowledgments
S.E.\ and H.F.\ gratefully acknowledge financial support 
by the Deutsche Forschungsgemeinschaft through the SFB 652.

\bibliographystyle{apsrev4-1}

\begin{thebibliography}{10}%
\makeatletter
\providecommand \@ifxundefined [1]{%
 \ifx #1\undefined \expandafter \@firstoftwo
 \else \expandafter \@secondoftwo
\fi
}%
\providecommand \@ifnum [1]{%
 \ifnum #1\expandafter \@firstoftwo
 \else \expandafter \@secondoftwo
\fi
}%
\providecommand \enquote [1]{``#1''}%
\providecommand \bibnamefont  [1]{#1}%
\providecommand \bibfnamefont [1]{#1}%
\providecommand \citenamefont [1]{#1}%
\providecommand\href[0]{\@sanitize\@href}%
\providecommand\@href[1]{\endgroup\@@startlink{#1}\endgroup\@@href}%
\providecommand\@@href[1]{#1\@@endlink}%
\providecommand \@sanitize [0]{\begingroup\catcode`\&12\catcode`\#12\relax}%
\@ifxundefined \pdfoutput {\@firstoftwo}{%
 \@ifnum{\z@=\pdfoutput}{\@firstoftwo}{\@secondoftwo}%
}{%
 \providecommand\@@startlink[1]{\leavevmode}%
 \providecommand\@@endlink[0]{}%
}{%
 \providecommand\@@startlink[1]{%
  \leavevmode
  \pdfstartlink
   attr{/Border[0 0 1 ]/H/I/C[0 1 1]}%
   user{/Subtype/Link/A<</Type/Action/S/URI/URI(#1)>>}%
  \relax
 }%
 \providecommand\@@endlink[0]{\pdfendlink}%
}%
\providecommand \url  [0]{\begingroup\@sanitize \@url }%
\providecommand \@url [1]{\endgroup\@href {#1}{\urlprefix}}%
\providecommand \urlprefix [0]{URL }%
\providecommand \Eprint[0]{\href }%
\@ifxundefined \urlstyle {%
  \providecommand \doi [1]{doi:\discretionary{}{}{}#1}%
}{%
  \providecommand \doi [0]{doi:\discretionary{}{}{}\begingroup
  \urlstyle{rm}\Url }%
}%
\providecommand \doibase [0]{http://dx.doi.org/}%
\providecommand \Doi[1]{\href{\doibase#1}}%
\providecommand \bibAnnote [3]{%
  \BibitemShut{#1}%
  \begin{quotation}\noindent
    \textsc{Key:}\ #2\\\textsc{Annotation:}\ #3%
  \end{quotation}%
}%
\providecommand \bibAnnoteFile [2]{%
  \IfFileExists{#2}{\bibAnnote {#1} {#2} {\input{#2}}}{}%
}%
\providecommand \typeout [0]{\immediate \write \m@ne }%
\providecommand \selectlanguage [0]{\@gobble}%
\providecommand \bibinfo [0]{\@secondoftwo}%
\providecommand \bibfield [0]{\@secondoftwo}%
\providecommand \translation [1]{[#1]}%
\providecommand \BibitemOpen[0]{}%
\providecommand \bibitemStop [0]{}%
\providecommand \bibitemNoStop [0]{.\EOS\space}%
\providecommand \EOS [0]{\spacefactor3000\relax}%
\providecommand \BibitemShut [1]{\csname bibitem#1\endcsname}%
\bibitem{BDZ08}%
  \BibitemOpen
  \bibfield{author}{%
  \bibinfo {author} {\bibfnamefont{I.}~\bibnamefont{Bloch}}, \bibinfo {author}
  {\bibfnamefont{J.}~\bibnamefont{Dalibard}},\ and\ \bibinfo {author}
  {\bibfnamefont{W.}~\bibnamefont{Zwerger}},\ }%
  \bibfield{journal}{%
  \bibinfo {journal} {Rev.\ Mod.\ Phys.}\ }%
  \textbf{\bibinfo {volume} {80}},\ \bibinfo {pages} {885} (\bibinfo {year}
  {2008})%
  \bibAnnoteFile{NoStop}{BDZ08}%
\bibitem{GMEHB02}%
  \BibitemOpen
  \bibfield{author}{%
  \bibinfo {author} {\bibfnamefont{M.}~\bibnamefont{Greiner}}, \bibinfo
  {author} {\bibfnamefont{O.}~\bibnamefont{Mandel}}, \bibinfo {author}
  {\bibfnamefont{T.}~\bibnamefont{Esslinger}}, \bibinfo {author}
  {\bibfnamefont{T.~W.}\ \bibnamefont{H\"ansch}},\ and\ \bibinfo {author}
  {\bibfnamefont{I.}~\bibnamefont{Bloch}},\ }%
  \bibfield{journal}{%
  \bibinfo {journal} {Nature}\ }%
  \textbf{\bibinfo {volume} {415}},\ \bibinfo {pages} {39} (\bibinfo {year}
  {2002})%
  \bibAnnoteFile{NoStop}{GMEHB02}%
\bibitem{WBSHLB10}%
  \BibitemOpen
  \bibfield{author}{%
  \bibinfo {author} {\bibfnamefont{S.}~\bibnamefont{Will}}, \bibinfo {author}
  {\bibfnamefont{T.}~\bibnamefont{Best}}, \bibinfo {author}
  {\bibfnamefont{U.}~\bibnamefont{Schneider}}, \bibinfo {author}
  {\bibfnamefont{L.}~\bibnamefont{Hackerm\"{u}ller}}, \bibinfo {author}
  {\bibfnamefont{D.-S.}\ \bibnamefont{L\"{u}h\-mann}},\ and\ \bibinfo {author}
  {\bibfnamefont{I.}~\bibnamefont{Bloch}},\ }%
  \bibfield{journal}{%
  \Doi{10.1038/nature09036}{\bibinfo {journal} {Nature}}\ }%
  \textbf{\bibinfo {volume} {465}},\ \bibinfo {pages} {197} (\bibinfo {year}
  {2010})%
  \bibAnnoteFile{NoStop}{WBSHLB10}%
\bibitem{MHLDDN11}%
  \BibitemOpen
  \bibfield{author}{%
  \bibinfo {author} {\bibfnamefont{M.~J.}\ \bibnamefont{Mark}}, \bibinfo
  {author} {\bibfnamefont{E.}~\bibnamefont{Haller}}, \bibinfo {author}
  {\bibfnamefont{K.}~\bibnamefont{Lauber}}, \bibinfo {author}
  {\bibfnamefont{J.~G.}\ \bibnamefont{Danzl}}, \bibinfo {author}
  {\bibfnamefont{A.~J.}\ \bibnamefont{Daley}},\ and\ \bibinfo {author}
  {\bibfnamefont{H.-C.}\ \bibnamefont{N\"agerl}},\ }%
  \bibfield{journal}{%
  \Doi{10.1103/PhysRevLett.107.175301}{\bibinfo {journal} {Phys. Rev. Lett.}}\
  }%
  \textbf{\bibinfo {volume} {107}},\ \bibinfo {pages} {175301} (\bibinfo {year}
  {2011})%
  \bibAnnoteFile{NoStop}{MHLDDN11}%
\bibitem{KM98}%
  \BibitemOpen
  \bibfield{author}{%
  \bibinfo {author} {\bibfnamefont{T.~D.}\ \bibnamefont{K{\"u}hner}}\ and\
  \bibinfo {author} {\bibfnamefont{H.}~\bibnamefont{Monien}},\ }%
  \bibfield{journal}{%
  \bibinfo {journal} {Phys.\ Rev.\ B}\ }%
  \textbf{\bibinfo {volume} {58}},\ \bibinfo {pages} {R14741} (\bibinfo {year}
  {1998})%
  \bibAnnoteFile{NoStop}{KM98}%
\bibitem{KWM00}%
  \BibitemOpen
  \bibfield{author}{%
  \bibinfo {author} {\bibfnamefont{T.~D.}\ \bibnamefont{K{\"u}hner}}, \bibinfo
  {author} {\bibfnamefont{S.~R.}\ \bibnamefont{White}},\ and\ \bibinfo {author}
  {\bibfnamefont{H.}~\bibnamefont{Monien}},\ }%
  \bibfield{journal}{%
  \bibinfo {journal} {Phys.\ Rev.\ B}\ }%
  \textbf{\bibinfo {volume} {61}},\ \bibinfo {pages} {12474} (\bibinfo {year}
  {2000})%
  \bibAnnoteFile{NoStop}{KWM00}%
\bibitem{EFG11}%
  \BibitemOpen
  \bibfield{author}{%
  \bibinfo {author} {\bibfnamefont{S.}~\bibnamefont{Ejima}}, \bibinfo {author}
  {\bibfnamefont{H.}~\bibnamefont{Fehske}},\ and\ \bibinfo {author}
  {\bibfnamefont{F.}~\bibnamefont{Gebhard}},\ }%
  \bibfield{journal}{%
  \bibinfo {journal} {Europhys.\ Lett.}\ }%
  \textbf{\bibinfo {volume} {93}},\ \bibinfo {pages} {30002} (\bibinfo {year}
  {2011})%
  \bibAnnoteFile{NoStop}{EFG11}%
\bibitem{EFGMKAL12}%
  \BibitemOpen
  \bibfield{author}{%
  \bibinfo {author} {\bibfnamefont{S.} \bibnamefont{Ejima}}, \bibinfo {author}
  {\bibfnamefont{H.} \bibnamefont{Fehske}}, \bibinfo {author}
  {\bibfnamefont{F.} \bibnamefont{Gebhard}}, \bibinfo {author}
  {\bibfnamefont{K.} \bibnamefont{zu~M\"{u}nster}}, \bibinfo {author}
  {\bibfnamefont{M.} \bibnamefont{Knap}}, \bibinfo {author}
  {\bibfnamefont{E.} \bibnamefont{Arrigoni}},\ and\ \bibinfo {author}
  {\bibfnamefont{W.} \bibnamefont{von~der Linden}},\ }%
  \bibfield{journal}{%
  \Doi{10.1103/PhysRevA.85.053644}{\bibinfo {journal} {Phys.\ Rev.\ A}}\ }%
  \textbf{\bibinfo {volume} {85}},\ \bibinfo {pages} {053644} (\bibinfo {year}
  {2012})%
  \bibAnnoteFile{NoStop}{EFGMKAL12}%
\bibitem{Wh92}%
  \BibitemOpen
  \bibfield{author}{%
  \bibinfo {author} {\bibfnamefont{S.~R.}\ \bibnamefont{White}},\ }%
  \bibfield{journal}{%
  \bibinfo {journal} {Phys.\ Rev.\ Lett.}\ }%
  \textbf{\bibinfo {volume} {69}},\ \bibinfo {pages} {2863} (\bibinfo {year}
  {1992})%
  \bibAnnoteFile{NoStop}{Wh92}%
\bibitem{SS11}%
  \BibitemOpen
  \bibfield{author}{%
  \bibinfo {author} {\bibfnamefont{J.}~\bibnamefont{Silva-Valencia}}\ and\
  \bibinfo {author} {\bibfnamefont{A.~M.~C.}\ \bibnamefont{Souza}},\ }%
  \bibfield{journal}{%
  \Doi{10.1103/PhysRevA.84.065601}{\bibinfo {journal} {Phys.\ Rev.\ A}}\ }%
  \textbf{\bibinfo {volume} {84}},\ \bibinfo {pages} {065601} (\bibinfo {year}
  {2011})%
  \bibAnnoteFile{NoStop}{SS11}%
\bibitem{VM12}%
  \BibitemOpen
  \bibfield{author}{%
  \bibinfo {author} {\bibfnamefont {V.}~\bibnamefont {Varma}}\ and\ %
  \bibinfo {author} {\bibfnamefont {H.}~\bibnamefont {Monien}},\ }%
  \bibinfo {note} {e-print, arXiv:1211.5664.}%
  \bibAnnoteFile{NoStop}{VM12}%
\bibitem{So12}%
  \BibitemOpen
  \bibfield{author}{%
  \bibinfo {author} {\bibfnamefont{T.}~\bibnamefont{Sowi\ifmmode~\acute{n}\else
  \'{n}\fi{}ski}},\ }%
  \bibfield{journal}{%
  \Doi{10.1103/PhysRevA.85.065601}{\bibinfo {journal} {Phys.\ Rev.\ A}}\ }%
  \textbf{\bibinfo {volume} {85}},\ \bibinfo {pages} {065601} (\bibinfo {year}
  {2012})%
  \bibAnnoteFile{NoStop}{So12}%
\bibitem{SDMPD12}%
  \BibitemOpen
  \bibfield  {author} {\bibinfo {author} {\bibfnamefont {M.}~\bibnamefont
  {Singh}}, \bibinfo {author} {\bibfnamefont {A.}~\bibnamefont {Dhar}},
  \bibinfo {author} {\bibfnamefont {T.}~\bibnamefont {Mishra}}, \bibinfo
  {author} {\bibfnamefont {R.~V.}\ \bibnamefont {Pai}}, \ and\ \bibinfo
  {author} {\bibfnamefont {B.~P.}\ \bibnamefont {Das}},\ }\href {\doibase
  10.1103/PhysRevA.85.051604} {\bibfield  {journal} {\bibinfo  {journal} {Phys.
  Rev. A}\ }\textbf {\bibinfo {volume} {85}},\ \bibinfo {pages} {051604}
  (\bibinfo {year} {2012})}\BibitemShut {NoStop}%
\bibitem{SCDTL13}%
  \BibitemOpen
  \bibfield{author}{%
  \bibinfo {author} {\bibfnamefont{T.}~\bibnamefont{Sowi\'nski}}, \bibinfo
  {author} {\bibfnamefont{R.~W.}\ \bibnamefont{Chhajlany}}, \bibinfo {author}
  {\bibfnamefont{O.}~\bibnamefont{Dutta}}, \bibinfo {author}
  {\bibfnamefont{L.}~\bibnamefont{Tagliacozzo}},\ and\ \bibinfo {author}
  {\bibfnamefont{M.}~\bibnamefont{Lewenstein}},\ }%
  \bibinfo {note} {e-print, arXiv:1304.4835.}%
  \bibAnnoteFile{NoStop}{SCDTL13}%
\bibitem{Je02b}%
  \BibitemOpen
  \bibfield{author}{%
  \bibinfo {author} {\bibfnamefont{E.}~\bibnamefont{Jeckelmann}},\ }%
  \bibfield{journal}{%
  \bibinfo {journal} {Phys.\ Rev.\ B}\ }%
  \textbf{\bibinfo {volume} {66}},\ \bibinfo {pages} {045114} (\bibinfo {year}
  {2002})%
  \bibAnnoteFile{NoStop}{Je02b}%
\bibitem{EGN05}%
  \BibitemOpen
  \bibfield{author}{%
  \bibinfo {author} {\bibfnamefont{S.}~\bibnamefont{Ejima}}, \bibinfo {author}
  {\bibfnamefont{F.}~\bibnamefont{Gebhard}},\ and\ \bibinfo {author}
  {\bibfnamefont{S.}~\bibnamefont{Nishimoto}},\ }%
  \bibfield{journal}{%
  \bibinfo {journal} {Europhys.\ Lett.}\ }%
  \textbf{\bibinfo {volume} {70}},\ \bibinfo {pages} {492} (\bibinfo {year}
  {2005})%
  \bibAnnoteFile{NoStop}{EGN05}%
\bibitem{Gi03}%
  \BibitemOpen
  \bibfield{author}{%
  \bibinfo {author} {\bibfnamefont{T.}~\bibnamefont{Giamarchi}},\ }%
  \emph{\bibinfo {title} {Quantum Physics in One Dimension}}\ (\bibinfo
  {publisher} {Cla\-ren\-don Press},\ \bibinfo {address} {Oxford},\ \bibinfo {year}
  {2003})%
  \bibAnnoteFile{NoStop}{Gi03}%
\bibitem{CC04}%
  \BibitemOpen
  \bibfield{author}{%
  \bibinfo {author} {\bibfnamefont{P.}~\bibnamefont{Calabrese}}\ and\ \bibinfo
  {author} {\bibfnamefont{J.}~\bibnamefont{Cardy}},\ }%
  \bibfield{journal}{%
  \bibinfo {journal} {J.\ Stat.\ Mech.: Theory and Exp.}\ }%
  \textbf{\bibinfo {volume} {\hbox{\rm P06002}}} (\bibinfo {year} {2004})%
  \bibAnnoteFile{NoStop}{CC04}%
\bibitem{Ni11}%
  \BibitemOpen
  \bibfield{author}{%
  \bibinfo {author} {\bibfnamefont{S.}~\bibnamefont{Nishimoto}},\ }%
  \bibfield{journal}{%
  \bibinfo {journal} {Phys.\ Rev.\ B}\ }%
  \textbf{\bibinfo {volume} {84}},\ \bibinfo {pages} {195108} (\bibinfo {year}
  {2011})%
  \bibAnnoteFile{NoStop}{Ni11}%
\bibitem{SICSPK99}%
  \BibitemOpen
  \bibfield{author}{%
  \bibinfo {author} {\bibfnamefont{J.}~\bibnamefont{Stenger}}, \bibinfo
  {author} {\bibfnamefont{S.}~\bibnamefont{Inouye}}, \bibinfo {author}
  {\bibfnamefont{A.~P.}\ \bibnamefont{Chikkatur}}, \bibinfo {author}
  {\bibfnamefont{D.~M.}\ \bibnamefont{Stamper-Kurn}}, \bibinfo {author}
  {\bibfnamefont{D.~E.}\ \bibnamefont{Pritchard}},\ and\ \bibinfo {author}
  {\bibfnamefont{W.}~\bibnamefont{Ketterle}},\ }%
  \bibfield{journal}{%
  \bibinfo {journal} {Phys.\ Rev.\ Lett.}\ }%
  \textbf{\bibinfo {volume} {82}},\ \bibinfo {pages} {4569} (\bibinfo {year}
  {1999})%
  \bibAnnoteFile{NoStop}{SICSPK99}%
\bibitem{EGKPLPS09}%
  \BibitemOpen
  \bibfield{author}{%
  \bibinfo {author} {\bibfnamefont{P.~T.}\ \bibnamefont{Ernst}}, \bibinfo
  {author} {\bibfnamefont{S.}~\bibnamefont{G\"otze}}, \bibinfo {author}
  {\bibfnamefont{J.~S.}\ \bibnamefont{Krauser}}, \bibinfo {author}
  {\bibfnamefont{K.}~\bibnamefont{Pyka}}, \bibinfo {author}
  {\bibfnamefont{D.-S.}\ \bibnamefont{L\"uhmann}}, \bibinfo {author}
  {\bibfnamefont{D.}~\bibnamefont{Pfannkuche}},\ and\ \bibinfo {author}
  {\bibfnamefont{K.}~\bibnamefont{Sengstock}},\ }%
  \bibfield{journal}{%
  \bibinfo {journal} {Nature Physics}\ }%
  \textbf{\bibinfo {volume} {6}},\ \bibinfo {pages} {56} (\bibinfo {year}
  {2009})%
  \bibAnnoteFile{NoStop}{EGKPLPS09}%
\bibitem{EFG12}%
  \BibitemOpen
  \bibfield{author}{%
  \bibinfo {author} {\bibfnamefont{S.}~\bibnamefont{Ejima}}, \bibinfo {author}
  {\bibfnamefont{H.}~\bibnamefont{Fehske}},\ and\ \bibinfo {author}
  {\bibfnamefont{F.}~\bibnamefont{Gebhard}},\ }%
  \bibfield{journal}{%
  \bibinfo {journal} {J.\ Phys.\ Conf.\ Ser.}\ }%
  \textbf{\bibinfo {volume} {391}},\ \bibinfo {pages} {012031} (\bibinfo {year}
  {2012})%
  \bibAnnoteFile{NoStop}{EFG12}%
\bibitem{NGJ04}%
  \BibitemOpen
  \bibfield  {author} {\bibinfo {author} {\bibfnamefont {S.}~\bibnamefont
  {Nishimoto}}, \bibinfo {author} {\bibfnamefont {F.}~\bibnamefont {Gebhard}},
  \ and\ \bibinfo {author} {\bibfnamefont {E.}~\bibnamefont {Jeckelmann}},\
  } {\bibfield  {journal} {\bibinfo  {journal} {J.\ Phys.\
  Condens.\ Matter}\ }\textbf {\bibinfo {volume} {16}},\ \bibinfo {pages}
  {7063} (\bibinfo {year} {2004})}\BibitemShut {NoStop}%
\bibitem{MuensterMSc} K.\ zu M\"unster, M.Sc.\ thesis, 
University of Marburg (2012).
\end{thebibliography}

\end{document}